# Photovoltaic grid-forming control strategy investigation using Hardware-in-the-Loop Experiments

*Somesh Bhattacharya[1*], Chrysanthos Charalambous[2], Anja Banjac[3], Zoran Miletic[3], Thomas Strasser[3], Brian Azzopardi[4,5], Christina Papadimitriou[6], Venizelos Efthymiou[2], Alexis Polycarpou[7]*

[1] Industrial Electric Power Conversion, Faculty of Engineering, The University of Malta, Msida, Malta
[2] University of Cyprus, Nicosia, Cyprus
[3] Austrian Institute of Technology, Vienna, Austria
[4] MCAST Energy Research Group, Institute of Engineering and Transport,
Malta College of Arts, Science and Technology (MCAST), Corradino Hill, Paola PLA9032, Malta
[5] The Foundation for Innovation and Research – Malta,
65 Design Centre Level 2, Tower Road, Birkirkara BKR4012, Malta
[6] Eindhoven University of Technology, 5612 AZ Eindhoven, The Netherlands
[7] Frederick University, Limassol, Cyprus

*somesh.bhattacharya@um.edu.mt

## Abstract

The frequency stability of a power system is of paramount importance, as a fast frequency swings in the system can lead to oscillatory instability, and thereby blackouts. A grid-connected microgrid, that can operate in the islanded mode can also possess such deteriorating effect due to the higher share of converter-based sources. In this paper, a coordinated frequency control within a distribution network is discussed, with a higher share of Photovoltaics (PV). The main objective of this paper is to test the grid-forming capabilities of PVs, without the requirement of an energy storage in the network. The tests were carried out with the help of the Typhoon Hardware-in-the-loop (HIL) platform using a real Cypriot network feeder. The real-time results confirm the efficacy of the PV as a grid-forming inverter, provided it has sufficient input (irradiance) to provide for the loads within the system of interest. The grid-forming PV also possesses the capability of reconnection with the utility grid through a synchronizer switch that requires minimal communication, makes the overall control independent of any other power source, subject to certain irradiance and loading conditions.

## 1 Introduction

For researchers and power system operators, frequency regulation from the perspective of low voltage distribution networks still presents a difficulty. With more and more PVs and batteries joining the grid and providing grid-related services like voltage and frequency regulation, active distribution networks are rapidly becoming a reality. Small islands like Malta and Cyprus have different power networks from those on the mainland. The primary distinction is the type of fuel utilized by conventional synchronous generators, which is often gas oil or heavy fuel oil. Due to the high levels of solar radiation, renewable energy, particularly photovoltaics (PVs), is widely used on both Islands, primarily as rooftop plants due to space restrictions. PVs, on the other hand, are inverter-based power sources and do not benefit from inertial support from nature. This feature is detrimental to the power system's dynamic stability, particularly when it comes to islands that aren't or are only loosely connected with other electrical networks [3]–[5].

It can cause an unacceptable rate of change of frequency (RoCoF), which can cause problems with the quality of the power, generator trips, unserved loads, and ultimately the complete failure of the system [6].

Rooftop PV proliferation increases the need for improved control concepts. In order to get inertial response for a typical weak or non-interconnected grid under the protection of a tiny Island like Malta or Cyprus, the major goal of this work is to harness virtual inertia from the PVs. The choice of which PV to take part in frequency response must be made using an approach that keeps the controls operational for all participating PVs. On theory, the largest capacity PV should be chosen because it should offer the most support; but, when all the PVs have about the same capacity (especially in an island scenario), making the decision is more challenging and there are also technological limitations [7], [8]. Therefore, a helicopter view is required to decide, for the set of PVs to work in the grid forming mode, and the rest in the grid following mode.

The main purpose of this work is to realize the PV in the distribution network to work in the grid-forming mode, especially once they have entered the islanded mode of operation. The advantages of PV working in the grid-forming mode are manifold, such as lower frequency transients during a load change (e.g. Fig. 1 where a load change is simulated and frequency response is observed), and less stress on the battery bank because of the active and the reactive power sharing. It is well known that in a distribution network, PVs are unable to engage in any market-based activities, and their owners only receive the installation subsidies. As a result, de-rating the PV to some extent will not have a negative effect on the owner. As a result, a cascaded switching-based controller is created for the PV in the Typhoon HIL environment and experiments are conducted on the condensed Cypriot network in order to achieve the goal of grid-forming PV control. The network consists of two aggregated PVs and a utility scale battery, and the latter works in the droop control mode, however, the inertial control isn't embedded as part of the battery controls. To ensure a smooth transition between the grid connected and the islanded modes, a synchronizer is also formulated which ensures a quick transition without any appreciable transients back to the grid-connected mode.

This document is organised as follows: Section 2 describes the methodology and briefly outlines the performed experiments, while section 3 presents the results of this work. Finally, section 4 summarises the conclusions of this study.

## 2. Methodology

The real-time simulation tests were performed through the HIL framework hosted at the AIT Austrian Institute of Technology premises. The test system considered for the real-time simulations is a single feeder of the distribution grid of Cyprus as presented in Fig. 2.

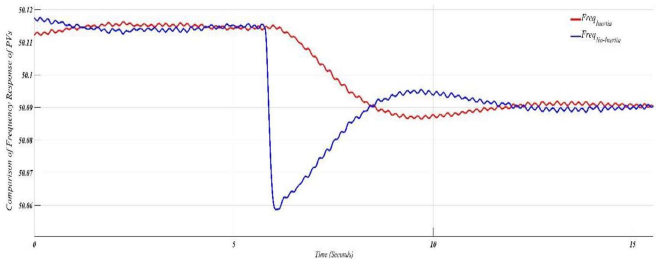

Fig. 1: Comparison of Responses of PV with and without inertial control.

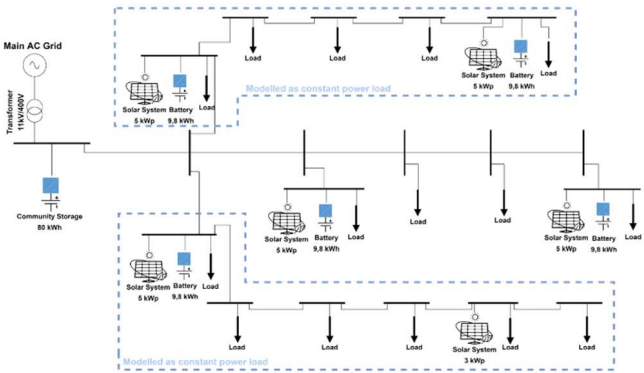

Fig. 2: Real feeder from the distribution grid of Cyprus.

The selected feeder composed from several small scale photovoltaic systems (<10 kWp), residential battery energy storage systems, residential loads, and a community battery energy storage system. This network is located on the outskirts of the province of Nicosia and is connected to a distribution substation of the distribution system of Cyprus. The feeder is connected to the medium voltage network through an 11kV / 400V transformer.

The community battery energy storage system is installed in the main substation and its technical specifications are shown in Table 1.

In addition to the community storage system, on this feeder there are other 4 residential battery energy storage systems installed. In the specific houses in which the batteries are installed, there is a photovoltaic system installed as well. The technical specifications of residential storage systems are presented in Table 2.

### 2.1 Test System Configuration and Equipment

As it was mentioned earlier in this paper, the tests are performed on a Cypriot distribution network with real data with two aggregated PVs. The term aggregated PVs comes from the fact that as there are several PVs of smaller capacities, in between 5kW and 9kW, the inverter controls are aggregated into a single controller for the ease of implementation, and to implicate lesser burden on the single arm processor of the HIL 604. The aggregated PVs are culminated into two PVs 22kW each. The Power-Voltage (P-V) and Current-Voltage (I-V) curves for the same can be seen in Fig. 3. The community battery energy storage, that is closest to the PCC is rated at 90kWh, that is connected to a VSI with 35kW nominal power rating.

Table 1: Technical specifications of the community storage system.

| Battery inverter | |
|---|---|
| Power rate | 30 KW / 30 KVar |
| Efficiency | 97% |
| Relative humidity range | 5% to 95% |
| Nominal grid voltage | 3PH: 400VAC ± 10% |
| Communication with battery | Modbus TCP/IP |
| Ingress Protection (IP) rate | IP21 |
| DC input voltage range | 100V – 500V / 360V |
| Installation | Indoor |
| **Battery Unit** | |
| Technology | Lithium Nickel Manganese Cobalt Oxide |
| Voltage Range | 384V to 498V |
| Relative humidity range | 5% to 95% |
| Installed battery unit capacity | Nominal: 80 kWh Usable : 50 kWh |
| Depth of Discharge (DoD) | 95% |
| Communication with the inverter | CAN bus |
| Ingress protection (IP) rate | IP21 |
| Operating Temperature Range | 5°C to 35°C |
| Installation | Indoor |

*Table 2: Technical specifications of the residential storage system.*

| Battery inverter | |
|---|---|
| Power rate/peak | 2.5 KW / 2.65KW |
| Round-trip efficiency | 94% |
| Relative humidity range | 0% to 100% |
| Communication with battery | CAN bus |
| Ingress Protection (IP) rate | IP65 |
| DC input voltage range / rated | 100V – 500V / 360V |
| Installation | Indoor or Outdoor |
| **Battery Unit** | |
| Technology | Lithium Iron Phosphate |
| Max. charge/discharge power | 5 KW |
| Relative humidity range | 5% to 95% |
| Installed battery unit capacity | Nominal: 9.8 kWh, Usable : 9.3 kWh |
| Depth of Discharge (DoD) | 95% |
| Communication with the inverter | CAN bus |
| Ingress protection (IP) rate | IP55 |
| Operating Temperature Range | -10°C to 45°C |
| Installation | Indoor or Outdoor |

In this section, the switching control strategy is covered in more detail. The design of the PV control allows it to operate in both current control mode (CCM) and voltage control mode (VCM). Once the PV is operating as a VCM in islanded mode, its VSG potential becomes apparent. Achieved while operating under CCM is standard MPPT operation.

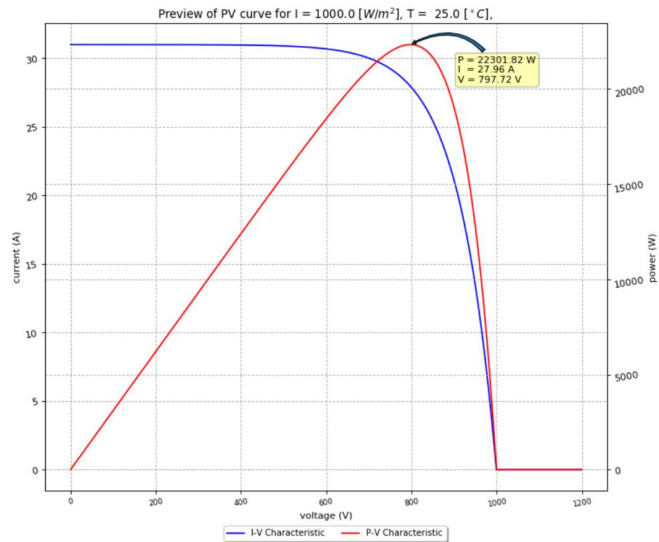

Figure 3: P-V and I-V curves for the photovoltaic systems.

The control approach for the PV as a grid-forming inverter, as seen in Fig. 4 is modelled as a switching controller. In the figure, an LC filter is used for the attenuation of the harmonic order frequency components. The output current and the voltage is sensed as $I_{i\_abc}$ and $V_{o\_abc}$ respectively. The grid currents and the voltages are sensed as $I_{o\_abc}$ and $V_{g\_abc}$ respectively. The DSOGI PLL [7] then is used for the estimation of the positive and negative phase angles ($\Theta^+_{PLL}$ and $\Theta^-_{PLL}$), along with the system frequency, ω.

By creating the frequency and voltage references with the help of virtual inertia and droop-based voltage control, the grid-forming control is accomplished. The latter control generates the voltage reference based on the loads' need for reactive power. This control is similar with the automatic voltage regulator (AVR) of the conventional synchronous generator. The virtual inertia control, and the voltage droop control as have been implemented using the typhoon HIL software can be described using the equations (3) and (4) as follow:

$$\omega_{ref} = \omega_{nom} - \frac{K_w}{T_w * S + 1}(P_{avg} - P_{ref}) \quad (3)$$

$$V_{ref} = V_{nom} - n_q(Q_{avg} - Q_{ref}) \quad (4)$$

In the equations above, the parameters $\omega_{ref}$ and $\omega_{nom}$ are the generated frequency reference, and the nominal frequency set at 100π radians per seconds (or 1 p.u.) respectively. The reference and the calculated average powers are $P_{ref}$ and $P_{avg}$. $K_w$ is the droop coefficient, while the time constant $T_w$ is H/D, where D is the damping coefficient. The inner voltage and current controllers are based on virtual admittance and the decoupled Synchronous reference frame (SRF) based controllers. The firing pulses are thus generated through the SRF control. The modulation index, $m_{abc}$ is then fed to the VSI. EN (1/0) in the figure is the *enable* option for the inverter which can be triggered through the supervisory control and data acquisition system (SCADA).

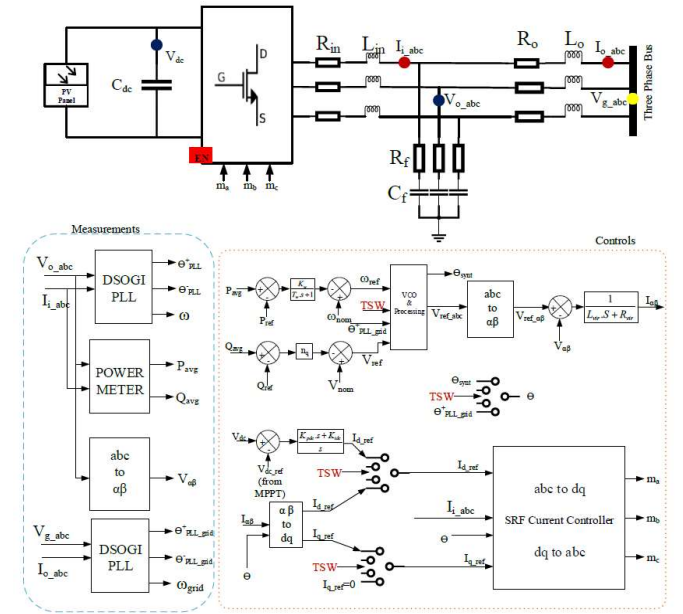

Fig. 4: Control strategy for the PV system as a grid forming inverter.

# 3 Results

This section summarizes the results of the work carried out during this project. The results are divided and presented into two different subsections depending the operation mode of the system. The grid-connected to islanded operation mode, and the islanded to grid-connected mode.

*3.1 Case Study 1- Grid-connected to islanded operation*

This sub-section describes the operation of the network, entering the islanded mode from the grid-connected operation. The responses of the PVs and the batteries are presented in Fig. 5 and Fig. 6 respectively.

Fig. 4 and Fig. 5 show that the batteries are initially set to be operating with Pref equal to 0, while the PVs operate at their MPPT. However, when the network switches to the islanded mode, all the sources split the available power according to the droop coefficient, $K_w$. Additionally, even in the presence of a switching-based controller, it can be observed that there are no transients during the changeover. The frequency of the PV systems is presented in Fig. 7.

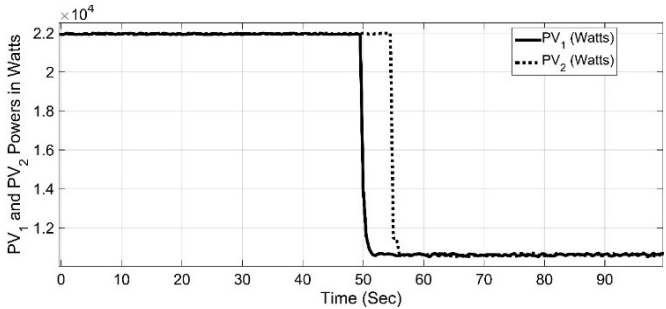

Fig. 5: Active power of the two aggregated PV systems.

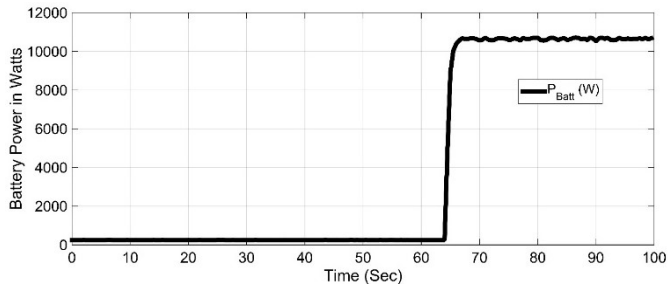

Fig. 6: Active power of the battery.

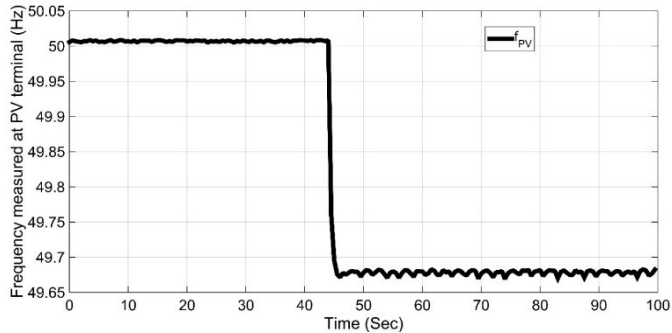

Fig. 7: PV system frequency during the grid-connected to islanded mode transition.

The instantaneous Battery and the PV voltages in the islanded mode, and the visible phase difference between the grid 'ϴ' (measured at the PCC) and the local VSI 'ϴ' (measured at the battery VSI point) is shown in Fig. 8. It can be seen in this figure that the voltage is 326V, i.e. the peak phase voltage, which is measured locally at the PV and the storage.

*3.2 Case Study 2- Islanded to Grid-Connected operation*

The Islanded to the grid-connected operation mode and the behaviour of the different components of the system, such as PVs and energy storage system, is presented in this sub-section. The active power of the battery energy storage system during the switching from islanded to grid-connected mode is presented in Fig. 9.

Furthermore, Fig. 10 presents the active power of the two aggregate PV systems. From this plot it can be observed that once the system switch from the islanded to grid-connected mode, the active power of the two PV systems reach again the MPPT position.

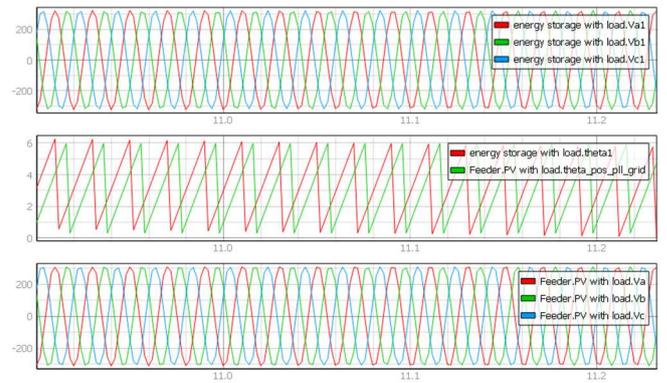

Fig. 8: Grid connected to Islanded mode Instantaneous Phase Voltages Battery and PV, and the phase difference.

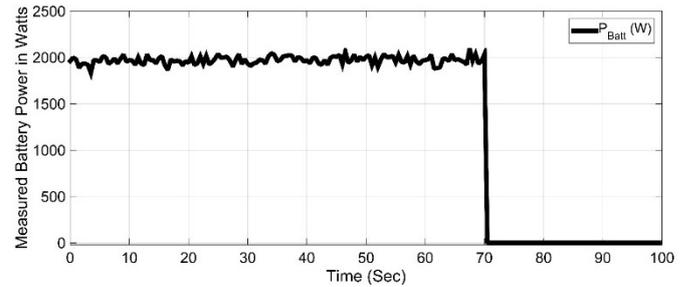

Fig. 9: Active power of the battery.

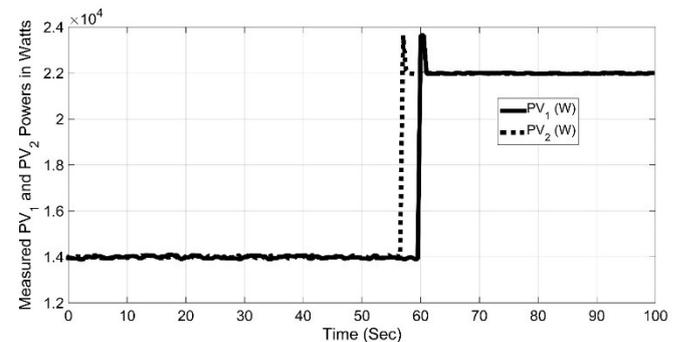

Fig. 10: Active power of the two aggregated PV systems.

Finally, minor transients are observed in the PV powers during the aforementioned transition and the battery also retains its initial '$P_{ref}=0$' position. The plots for the frequency, and the instantaneous measurements (voltage and phase angles) are shown in Fig. 11 and Fig. 12 respectively.

## 4 Conclusion

The present work focused on the grid-forming capability of the PV, that can seamlessly work in both grid-connected and islanded modes of operation. The designed grid-forming based switching controls for the PV were tested on a single feeder of the distribution grid of Cyprus using Typhoon HIL 604 based real-time simulations. The main findings of the simulations and tests carried out during this work are as follow

1. The system reliability is increased when all PVs are operating in grid-forming mode, particularly when operating on an island mode, as frequency and voltage variations from nominal values are low and all sources can adapt to network load fluctuations.
2. Virtual inertia-based design of the outer controller for the PV helps in transient-free load change, which helps in maintaining the RoCoF to the minimum value.

The studies and the analysis implemented under the current project investigates only the grid-forming VSI for PV, which forms the basis for further studies. Further studies can include the following approaches:

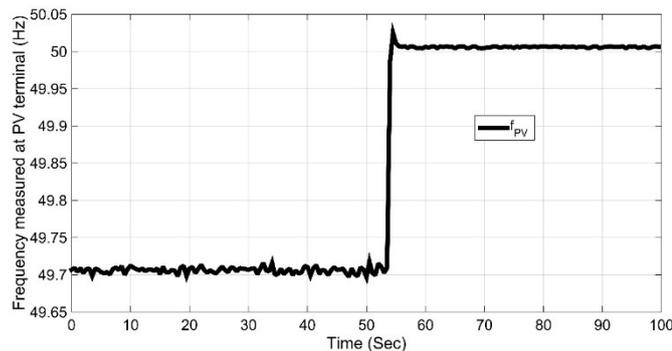

Fig. 11: PV system frequency during the islanded to grid-connected mode transition.

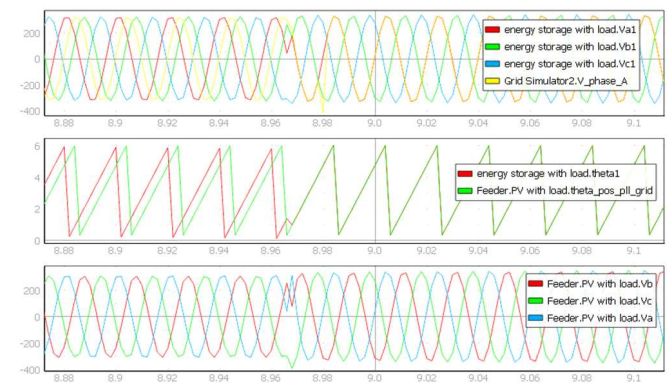

Fig. 12: Islanded to Grid-connected mode Instantaneous Phase Voltages Battery and PV, and the phase difference.

1. Developing a holistic control for the grid-forming PV, that can provide both frequency and voltage regulation services, which essentially includes low voltage ride through capability in both modes of operation.
2. Another approach can be the investigation and evaluation of the location of the PV systems in order to find the optimum location for the grid-forming PV in a broad distribution network.

## 5 Acknowledgements

This work was supported in part by the European Commission H2020 ERIGRID2.0 Project under Grant 870620 and TWINNING Networking for Excellence in Electric Mobility Operations (NEEMO) Project under Grant 857484.